\documentclass[10pt, conference]{IEEEtran}

\usepackage{cite}
\usepackage{amssymb}
\usepackage{mathtools}
\usepackage{amsmath}
\usepackage{graphicx}
\graphicspath{ {./image/} }
\usepackage{color}
\ifCLASSOPTIONcompsoc
    \usepackage[caption=false, font=normalsize, labelfont=sf, textfont=sf]{subfig}
\else
\usepackage[caption=false, font=footnotesize]{subfig}
\fi
\newcommand{\fakeparagraph}[1]{\vspace{.1mm}\noindent\textbf{#1}}
\newcommand{\fakepar}[1]{\fakeparagraph{#1}}

\usepackage[binary-units = true, range-phrase=--, per-mode=symbol]{siunitx}
\DeclareSIUnit{\decibelm}{dBm}
\DeclareSIUnit{\Joule}{Joule}
\DeclareSIUnit{\Sample}{S}
\usepackage{xcolor}

\begin{document}

\title{Edge Intelligence in Softwarized 6G: Deep 
Learning-enabled Network Traffic Predictions}
\author{
    \IEEEauthorblockN{Shah~Zeb\IEEEauthorrefmark{1},
    Muhammad~Ahmad~Rathore\IEEEauthorrefmark{2},
    Aamir~Mahmood\IEEEauthorrefmark{3},  Syed~Ali~Hassan\IEEEauthorrefmark{1}, JongWon~Kim\IEEEauthorrefmark{2},\\
    and~Mikael~Gidlund\IEEEauthorrefmark{3}}
\IEEEauthorblockA{\IEEEauthorrefmark{1}School of Electrical Engineering \& Computer Science (SEECS),\\ National University of Sciences \& Technology (NUST), Pakistan.}    
\IEEEauthorblockA{\IEEEauthorrefmark{2}School of Electrical Engineering \& Computer Science,\\ Gwangju Institute of Science \& Technology (GIST), Gwangju 61005, South Korea}    
\IEEEauthorblockA{\IEEEauthorrefmark{3}Department of Information Systems \& Technology, Mid Sweden University, Sweden.}

Email: \IEEEauthorrefmark{1}\{szeb.dphd19seecs,ali.hassan\}@seecs.edu.pk,
\IEEEauthorrefmark{2}\{ahmadrathore,jongwon\}@gist.ac.kr,\IEEEauthorrefmark{3}\{firstname.lastname\}@miun.se. 
\vspace{-10pt}
}

\maketitle

\begin{abstract}
The 6G vision is envisaged to enable agile network expansion and rapid deployment of new on-demand microservices (e.g., visibility services for data traffic management, mobile edge computing services) closer to the network's edge IoT devices.
However, providing one of the critical features of network visibility services, i.e., data flow prediction in the network, is challenging at the edge devices within a dynamic cloud-native environment as the traffic flow characteristics are random and sporadic.
To provide the AI-native services for the 6G vision, we propose a novel edge-native framework to provide an intelligent prognosis technique for data traffic management in this paper.
The prognosis model uses long short-term memory (LSTM)-based encoder-decoder deep learning, which we train on real time-series multivariate data records collected from the edge $\mu$-boxes of a selected testbed network.
Our result accurately predicts the statistical characteristics of data traffic and verifies the trained model against the ground truth observations. Moreover, we validate our novel framework with two performance metrics for each feature of the multivariate data.

\end{abstract}

\begin{IEEEkeywords}
6G, network traffic flow, forecasting, deep learning, cloud-native deployments, edge computing
\end{IEEEkeywords}
\IEEEpeerreviewmaketitle

\section{Introduction}
The successful commercialization of 5G networks paves the way for discussion on the next evolution towards 6G networks and defines it's vision and requirements~\cite{saad2019vision}.
While 5G architecture is built on service-based architecture (SBA)~\cite{Aamir_IIoT}, the vision of 6G hyper-flexible architecture revolves around the state-of-the-art artificial intelligence (AI)-native design that brings the intelligent decision-making abilities in futuristic applications of digital society, such as digital twin-enabled self-evolving innovative industries~\cite{Roadmap,zeb2021industrial}. 
The Third Generation Partnership Project (3GPP) is reportedly shifting towards new AI-inspired models to monitor and enhance the SBA performance~\cite{ref}.

The critical functional component to the enhanced core network will be the seamless convergence of AI models, software-defined network (SDN) and network function virtualization (NFV)-enabled communication networks, and edge/hybrid cloud-native computing architecture~\cite{xiao2020toward,zeb2020}. Similarly, virtualization and containerization are an integral part of cloud-native computing infrastructure, which lies in the domain of software development and IT operations (DevOps)~\cite{waseem2020systematic}. By adopting DevOps-based design strategies, telecom companies can successfully provide and break down purpose-built hardware services (i.e., edge computation) and software-based solutions into real microservices (i.e., orchestrate application deployments)~\cite{kube5g}. These microservices then move towards the edge of a network to satisfy key performance indicators (KPIs), e.g., data rates, network latency, energy efficiency~\cite{SDNandCloudflow}. Collectively, these new features contribute towards the intelligence in visibility services (i.e., monitoring KPIs) to futuristic innovative network operations and management systems for accurate network traffic flows forecasting. Besides, during the network operations, the data traffic flow having a time-series (TS) data nature behaves non-linearly with aperiodic characteristics in an increasingly dynamic and complex network environment~\cite{timeseries}. Similarly, the integration of popular Internet-of-thing (IoT) technology with a de-facto cloud/edge computing model increases the importance of visibility services in inferring future traffic behavior from past traffic for providing enhanced quality-of-service (QoS) and quality-of-experience (QoE)~\cite{IoTQos}.

Despite significant potential usage of AI and other technologies, their widespread deployment is yet to be seen for predicting network traffic as there are many challenges to their comprehensive adoption in networked systems. 

\fakepar{Insufficient resources for TS data}: Storage and computational resources needed for executing AI algorithms over network data flow near the edge network are limited and insufficient~\cite{Roadmap}. Meanwhile, as a high number of IoT devices connect to access networks, traffic volume unprecedentedly grows.

\fakepar{Need for large high-quality labeled data}: 
Machine learning (ML) algorithms need a large amount of labeled data for model training and learning, while most of the data from traffic flow at network points are unlabelled TS raw data that needs to be processed~\cite{timeseries}. Moreover, software and hardware network configurations, highly random sporadic geographic events, network infrastructure distribution, and other phenomena can lead to an abrupt change in traffic flows, affecting the quality of labeled data for the AI model construction.

\fakepar{Optimized network architecture for AI}: The design of existing network infrastructure lags behind the support for AI-inspired applications and services. Network resources can be drained off with the deployment of AI-based solutions~\cite{tang2021computing}. Therefore, the current networking infrastructure needs to evolve and adapt cloud-native AI deployment strategies to provide a balanced support for both the AI-based microservices functions and other diverse network functions.

One potential solution to approaching the preceding challenges is adopting \textit{edge intelligence or edge-native AI framework design} that integrates AI models, state-of-the-art communication networks, and cloud-native edge computing design to accurately predict traffic flows.
In this work, we propose a novel idea of an edge intelligence method for analyzing and predicting network data traffic at edge devices according to the emerging 6G vision of softwarized networks. For this purpose, we utilize the testbeds resources of cloud-native enabled \textit{OpenFlow over Trans-Eurasia Information Network (OF@TIEN++) Playground}, a multisite cloud connecting ten sites in nine Asian countries over TEIN network~\cite{rathore2020maintaining}. 
The main contributions of this paper are as follows.
\begin{itemize}
\item We present the novel method for predicting statistical characteristics of data traffic inflow at the edge devices of the cloud-native enabled network using a deep learning (DL) method.
\item For this purpose, we collect the time series raw traffic flow data at the edge of the network, which is sent to the visibility center for storage and processing.
\item We orchestrate the Kubeflow deployment at the orchestration center, which is used to develop and train the DL model for the prognosis.
\item We train the model on collected TS data and predict the statistical properties of data traffic. Moreover, we analyze the developed model's performance in terms of root-mean-square error (RMSE) and coefficient of determination (R\textsuperscript{2}) metrics.
\end{itemize}
The rest of this article is structured as follows. Section~\ref{sec:sysModel} presents the experimental model. Section~\ref{sec:performance} provides the discussion on designing Kubeflow-based deployment of a learning service. Section~\ref{sec:Results} presents the system validation and results. Finally, Section~\ref{sec:Conclusion} concludes this paper.
\section{Experimental Model Description}
\label{sec:sysModel}
This section provides the details of designing the Kubernetes (K8s)-based edge cluster with GIST playground control center, and packet tracing and flows summarization for dataset collection.
\subsection{OF@TEIN Playground Overview}
\label{sec:k8sedge}
\begin{figure}
\centering
\includegraphics[width=1\linewidth]{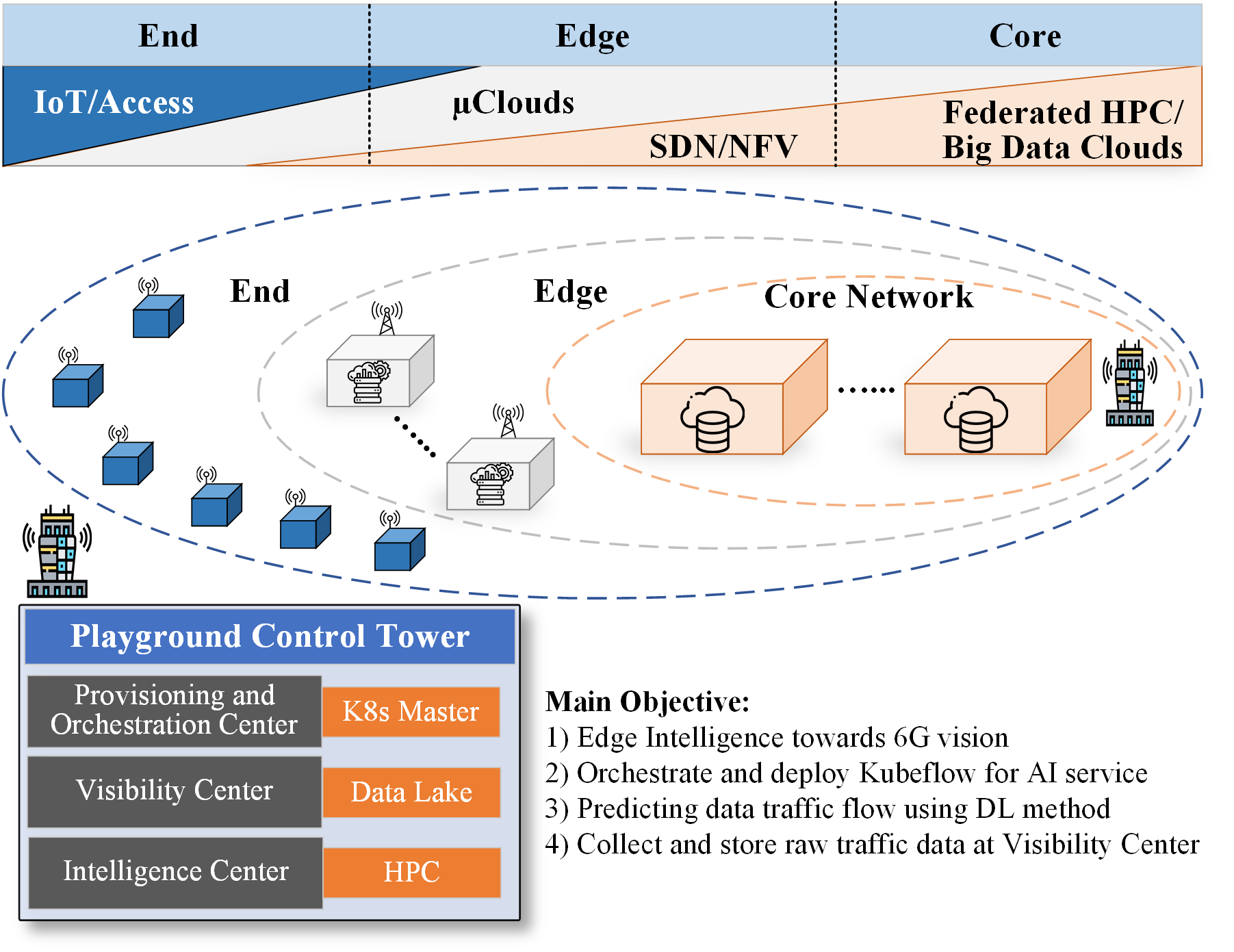}
	\caption{Illustration of K8s-based edge cluster over OF@TEIN playground.}
	\label{fig:PGInfratructure}
\vspace{-10pt}	
\end{figure}

The SDN-enabled multi-site clouds of OF@TEIN playground (OPG) interconnect multiple National Research and Education Network (NREN) of partner countries, enabling miniaturized academic experiments. 
Launched in 2020, it served as an \textit{Open Federated Playgrounds for AI-inspired SmartX Services} that has support for \textit{IoT-SDN/NFV-Cloud} functionalities.
Fig.~\ref{fig:PGInfratructure} shows the layered communication illustration of multi-site OPG infrastructure.
The playground (PG) within OPG supports the logical space in each centralized location/site called \textit{SmartX PG Tower} designated to develop, administer, and utilize the resources of distributed server-based hyper-converged cloud-native special boxes automatically. 
It maintains and dynamically distributes numerous physical and virtual resources for developers to execute research experiments and validate operational and development requirements in real-time.
Note that SmartX PG Tower leads the monitor and control functions of the network by installing and utilizing multiple operating centers, which are Provisioning and Orchestration (P+O), Visibility (V), and Intelligence (I).
\subsection{K8s Edge Cluster over OF@TEIN Playground}
\label{sec:k8sedgecluster}
In this work, we selected the P+O center of PG Tower at the GIST site to orchestrate and deploy an AI-based learning microservice using Kubeflow to forecast traffic flows based on the accumulated multivariate processed data set. We extract processed data from the strings of measured visibility flows collected from the SmartX MicroBox ($\mu$-box) intended for the data lake storage in the visibility center. We placed these $\mu$-boxes at the multi-site edge locations of the OPG, having computing-storage-networking resources to allow IoT-Cloud-SDN/NFV functionalities-based experiments.
To provide tenacious multi-access networked connectivity in each $\mu$-box, we enabled three network interfaces, i.e., two wired network interfaces and one wireless connectivity interface. Two network interfaces are assigned public Internet Protocols (IP)-based addresses and configured as control interface and data interface. Moreover, data from the connected IoT devices or data lakes are offloaded over-the-air to $\mu$-boxes.  
Afterward, we prepared and configured each $\mu$-box as K8s-orchestrated worker nodes to support cloud-native containerized functionalities, which are provisioned and managed from K8s master in the P+O center of GIST PG. Each $\mu$-box has SDN-coordinated unique/dedicated connectivity with other boxes supporting mesh-style networking and forming K8s edge clusters over OPG networked environment. A private IP addressing scheme is employed inside each $\mu$-box, which the K8s master manages for orchestrated pods and container communication.
\subsection{Network Traffic Data Set Collection}
\label{sec:dataset}
We use extended Berkeley Packet Filtering (eBPF)-enabled packet tracing tools such as IO Visor for measuring statistical summary of network data traffic. IO Visor-based packet tracing employs the eBPF core functionalities \cite{rathore2017comparing}, which enables in-kernel virtual machines (VMs) with byte-code tracing program execution. IO Visor has the main advantage to monitor and trace user and kernel events (through \textit{kprobe} and \textit{uprobe}), providing statistics in maps fetched on the points of interest \cite{gregg2016linux}. 
\textit{To collect network packet flows periodically}, we implement a data collection software that leverages eBPF and IO Visor to directly accumulate raw packets from the network interface of $\mu$-box and enables information compilation from each packet with a small number of CPU cycles (c.f~ Fig.~\ref{fig:AIservice}). At the same time, Apache-spark with Scala application program interface (API) is utilized to facilitate scalable, high-throughput, and fault-tolerant flows processing. The processed data flows generate five features containing a statistical summary of the network traffic packet. Later the generated multivariate data are pushed out using Spark stream and stored at MongoDB as a NoSQL database leveraging distributed messaging store of Apache Kafka.

\section{Kubeflow-based AI Service Design}
\label{sec:performance}
\begin{figure}
\centering
\includegraphics[width=1\linewidth]{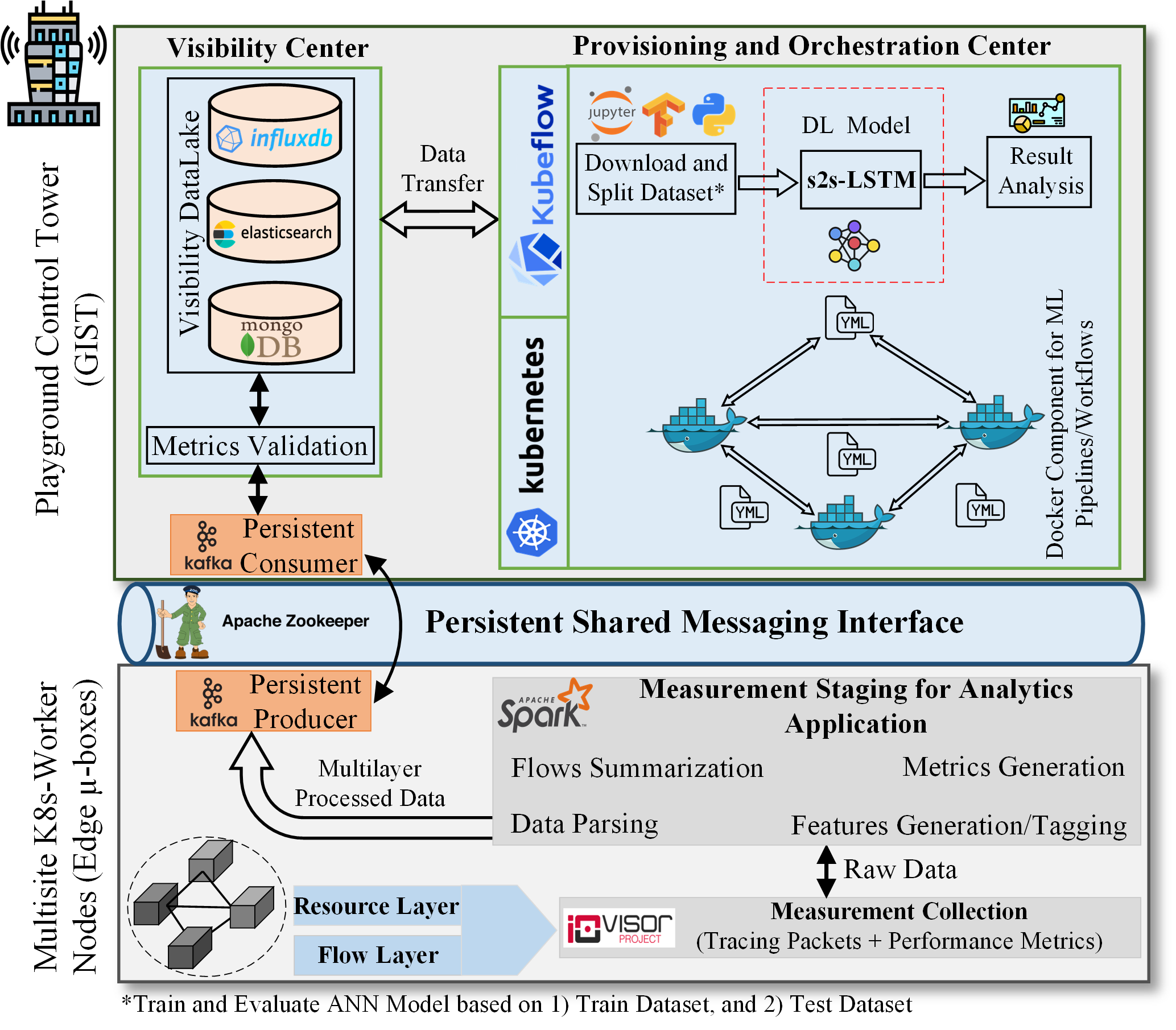}
	\caption{Design of Kubeflow-based AI service orchestration in K8s cluster of GIST site's P+O center.}
	\label{fig:AIservice}
	\vspace{-10pt}
\end{figure}
Kubeflow consists of toolsets that enable and inscribe numerous critical stages of the ML/DL development cycle, including preparing data, model learning, experimentation and tuning, and feature extractions/transformation. Moreover, Kubeflow leverages the benefits of the K8s cluster HPC capabilities for container orchestration and auto-scaling computing resources for ML/DL jobs/pipelines. Therefore, we deploy and orchestrate the Kubeflow using K8s Master at the P+O center of GIST PG to perform the high-performance data analytics (HPDA) by leveraging K8s cluster capabilities (c.f. Fig.~\ref{fig:AIservice}). It enables the platform for accurate data traffic prediction by applying the DL algorithm on collected TS data flow, explained in the subsequent sections. 
\subsection{DL-based Data Prediction Model}
\label{sec:DLframeworks}
\begin{figure}
\centering
\includegraphics[width=0.9\linewidth]{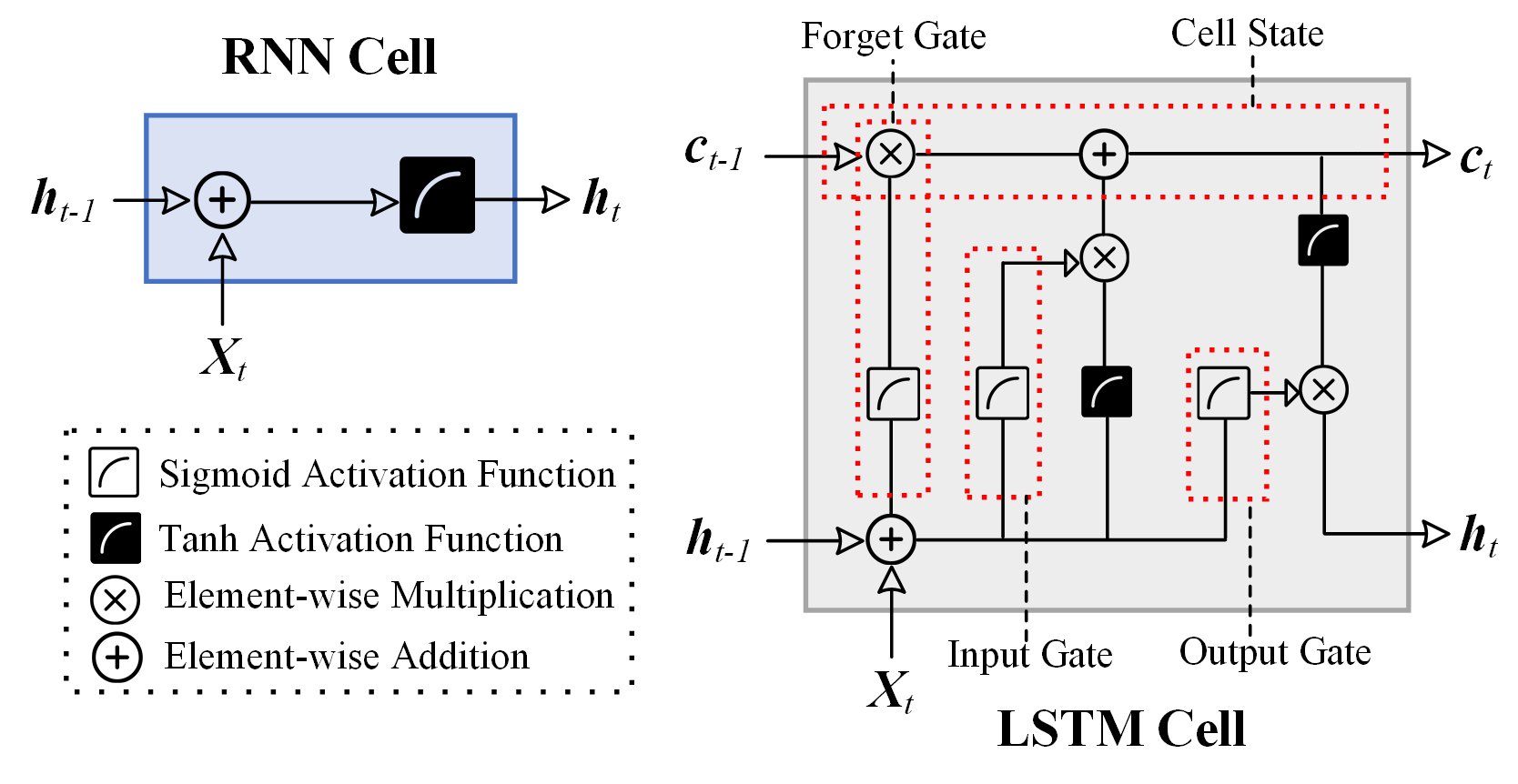}
	\caption{Folded representation of RNN and an LSTM unit cell.}
	\label{fig:RNNInfratructure}
	\vspace{-10pt}
\end{figure}
Traditional neural networks (NN) cannot utilize the information learned in the previous steps (past observations) to make the spatio-temporal learning on TS data and accurately predict traffic features.
Numerous recurrent NN (RNN) algorithms are developed for the prediction problems based on the unit RNN cell architecture (c.f.~Fig.~\ref{fig:RNNInfratructure}) due to their natural interpretation property of TS data analysis. These RNN algorithms allow data information to persists by connecting the previous informational state to the present task as an input. However, their performance suffers from the constraints in learning long-term dependencies/correlations of the TS data because of the vanishing gradient problem. In this paper, we use a sequence-to-sequence (s2s) deep learning model to predict based on long short-term memory (LSTM) neuron cells. In the following subsections, we first explain the LSTM cell and then discuss the encoder-decoder architecture for prediction based on the LSTM cell. 
\subsubsection{LSTM Cell}
An LSTM cell overcomes the RNNs vanishing gradient problem using back-propagation algorithms over time~\cite{LSTM}. The error derivatives for learning newly updated weights do not quickly vanish as they are distributed over sums and sent back in time, enabling LSTM units to learn and discover long-term correlated features over lengthy sequences in input multi-variate data.
As shown in Fig.~\ref{fig:RNNInfratructure}, an LSTM cell receives an input sequence vector $\mathbf{X}_t$ at current time $t$ which, together with the previous cell state $\mathbf{c}_t$ and hidden state $\mathbf{h}_t$, is used to trigger the different three gates by utilizing their activation processing units. Note that onward in the study, bold variable notation denotes a vector. A cell state of the LSTM unit can be considered as a memory unit, and its state can be read and modified through connected three gates. 
These LSTM unit gates are, 1) \textit{forget gate} ($\mathbf{f}_t$), which sets and decides what information to discard based on assigned condition, 2) \textit{input gate} ($\mathbf{i}_t$), which updates the memory cell state based on assigned conditions, and 3) \textit{output gate} ($\mathbf{o}_t$), which sets the output depending upon the input sequence and cell state with assigned conditions. 
The gates and cell updates of the LSTM unit at time $t$ can be formulated as 
\begin{equation*}\label{eq1}
{\mathbf{c}_t} = {\mathbf{f}_t} \odot {\mathbf{c}_{t - 1}} + {\mathbf{i}_t}\odot\tanh \left( {{\mathbf{w}_{hc}}\odot{\mathbf{h}_{t - 1}} + {\mathbf{w}_{xc}}\odot{\mathbf{X}_t} + {\mathbf{b}_c}} \right),
\end{equation*}
\begin{equation*}\label{eq2}
{\mathbf{i}_t} = \sigma \left( {{\mathbf{w}_{hi}}\odot{\mathbf{h}_{t - 1}} + {\mathbf{w}_{ci}}\odot{\mathbf{c}_{t - 1}} + {\mathbf{w}_{xi}}\odot{\mathbf{X}_t} + {\mathbf{b}_i}} \right),
\end{equation*}
\begin{equation*}\label{eq3}
{\mathbf{f}_t} = \sigma \left( {{\mathbf{w}_{hf}}\odot{\mathbf{h}_{t - 1}} + {\mathbf{w}_{cf}}\odot{\mathbf{c}_{t - 1}} + {\mathbf{w}_{xf}}\odot{\mathbf{X}_t} + {\mathbf{b}_f}} \right),
\end{equation*}
\begin{equation*}\label{eq4}
{\mathbf{o}_t} = \sigma \left( {{\mathbf{w}_{ho}}\odot{\mathbf{h}_{t - 1}} + {\mathbf{w}_{co}}\odot{\mathbf{c}_t} + {\mathbf{w}_{xo}}\odot{\mathbf{X}_t} + {\mathbf{b}_o}} \right),
\end{equation*}
\begin{equation*}\label{eq5}
{\mathbf{h}_t} = {\mathbf{o}_t}\tanh ({\mathbf{c}_t}),
\end{equation*}
where $\odot$ is the element-wise multiplication. Please note that each selected gate has a distinctly associated weight vector $\mathbf{w}$, and a bias vector $\mathbf{b}$, that are learned throughout the change of state and new information addition during processing in the training phase.
Moreover, each LSTM gate uses specific activation function (c.f.~Fig.~\ref{fig:RNNInfratructure}) for processing, e.g., sigmoid ($\sigma$) or hyperbolic tangent ($\mathrm{\tanh}$)~\cite[Sec.~3]{zhengtime}.
\begin{figure}
\centering
\includegraphics[width=0.82\linewidth]{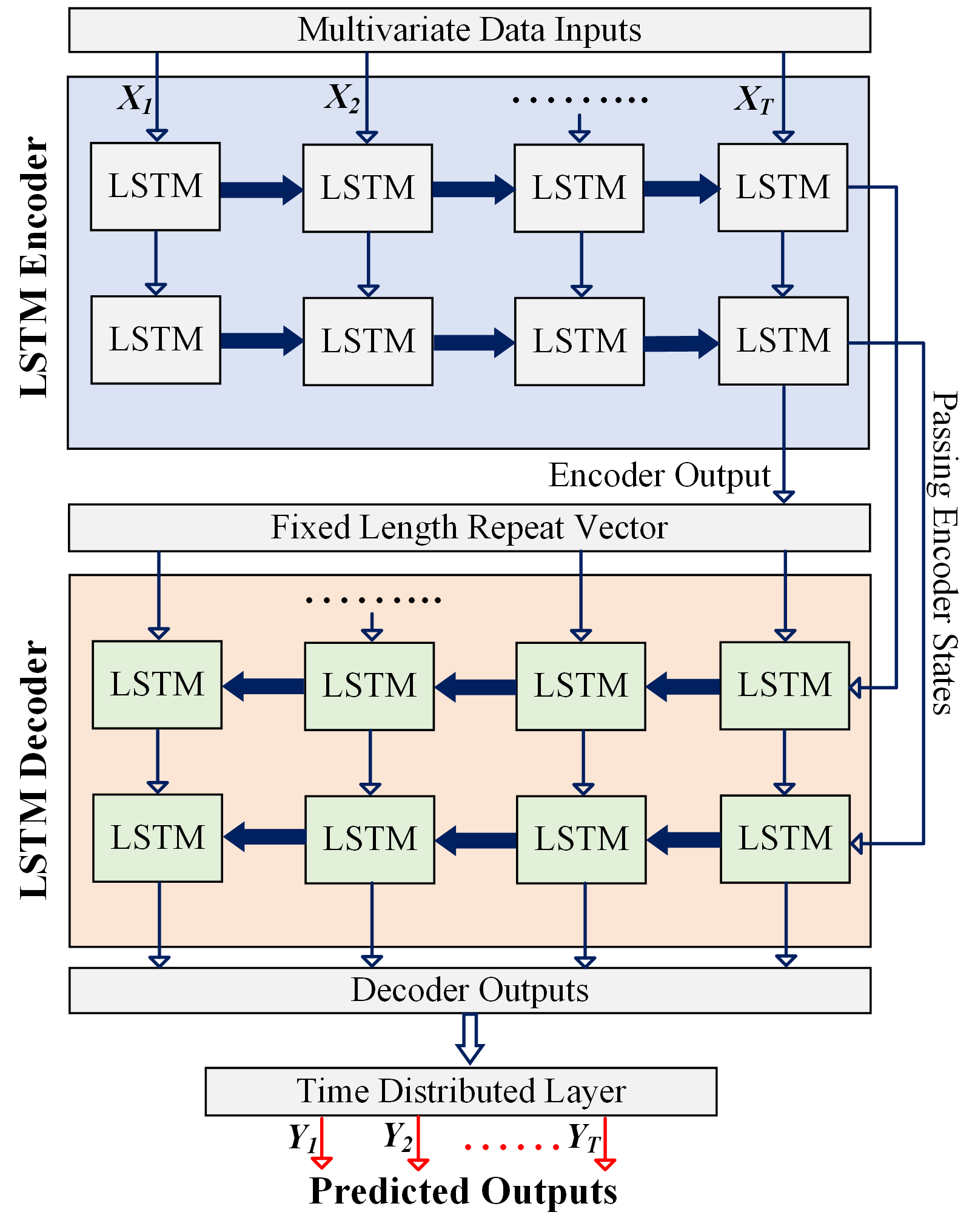}
	\caption{The schematic illustration shows the proposed DL-based prediction model with encoder-decoder architecture.}
	\label{fig:s2s}
	\vspace{-10pt}
\end{figure}
\subsubsection{LSTM cell-based Encoder-Decoder Model}
We used the single layer encoder-decoder architecture that employs the LSTM cells (c.f.~Fig.~\ref{fig:s2s}) in each layer to predict the vector set of output sequences of data traffic, $\mathcal{Y}_o=\{\mathbf{Y}_{t+1},\mathbf{Y}_{t+2},...,\mathbf{Y}_{t+f}\}$, based on the set of TS input sequences, $\mathcal{X}_i=\{\mathbf{X}_{t-T},\mathbf{X}_{t-T-1},...,\mathbf{X}_{t-1},\mathbf{X}_{t}\}$, which represents all the past observations of collected traffic. Note that $\mathbf{X}_{t}=\{x_{t,1}, x_{t,2},...,x_{t,m}\}$ represents the current time-dependent sequence vector containing the observation values for $m$ number of multivariate traffic features, $T$ is the lookback length of time, $f$ is the horizon window of future prediction and, $t,m,f,T \in \mathbb{N}$. 
The encoder produces the encoded temporal representation of current information sequences $\mathbf{X}_{t}$ through LSTM units in a single layer. The encoded output sequence vector is provided to the LSTM decoder through a repeat vector. Similarly, the encoder status of LSTM units is also simultaneously passed to the decoder units. Then, the decoder uses the cell state of the repeat vector as the initial temporal representation to reconstruct network data's target output, i.e., feature prediction. 
Now the final objective of the prediction model training problem is minimizing the output error $e_i$ for training samples in each $t$-th current time to find the optimized parameter space, $\Theta$, as,
\begin{equation}\label{eq6}
\mathop {\arg \min \: {e_i}}\limits_\Theta   = \sum\limits_{i = 1}^T {\sum\limits_{j = 1}^m {L_i^j} } \:,
\end{equation}
where $L_i^j$ is the selected Huber loss function for this study, and given as
\begin{equation}\label{eq7}
L_i^j\!\left( {X_i^j,Y_i^j} \right) \!\! = \!\! \left\{\!\! {\begin{array}{*{20}{c}}
{\frac{1}{2}{{\left( {X_i^j - Y_i^j} \right)}^2},}&{\forall \left| {X_i^j - Y_i^j} \right| \le \tau ,}\\
{\tau \left( {\left| {X_i^j - Y_i^j} \right| - \frac{1}{2}\tau } \right),}&{\mathrm{otherwise}.}
\end{array}} \right.
\end{equation}
In \eqref{eq6} and \eqref{eq7}, $\tau$ is the hyperparameter cut-off threshold to switch between two error functions (squared loss and absolute loss) which is 1 in our study,  $i,j \in \mathbb{N}$ represents the $j$-th feature observation value at each time-step $i$ in the input/predicted sequence of data, and $\Theta$ comprises of learned weights $\mathbf{w}$ and biases vectors $\mathbf{b}$ at each time-step. 
\section{Experimental Results Analysis}
\label{sec:Results}
{\renewcommand{\arraystretch}{1.1}
\begin{table}[t!]
\centering
	\caption{Device Specifications of Control Tower and $\mu$-boxes} 
	\scalebox{0.92}{
\begin{tabular}{|c|c|}
\hline
 \textbf{Device Type} & \textbf{Specifications} \\ \hline
 \begin{tabular}[c]{@{}c@{}} Visibility\\ Center\end{tabular} &
 \begin{tabular}[c]{@{}c@{}}Ubuntu 16.04.4 LTS OS, Intel Xeon{\textsuperscript \textregistered}CPU E5-2690 \\ V2@3.00GHz, 12x8 GB DDR3 Memory, 5.5 TB \\HDD, 4 network interfaces of 1 Gbits/sec (Gbps)  \end{tabular}  \\ \hline
 
   \begin{tabular}[c]{@{}c@{}} Provisioning \& \\ Orchestration \\ Center\end{tabular}  &  \begin{tabular}[c]{@{}c@{}}Ubuntu 18.04.2 LTS OS,SYS-E200-8D SuperServer\\ with Intel Xeon{\textsuperscript \textregistered} D-1528 with 6 Cores @1.90 GHz,\\ 32 GB RAM, 480 GB Intel SSD DC S3500   \end{tabular} \\ \hline
   
   \begin{tabular}[c]{@{}c@{}} Intelligence \\ Center\end{tabular}  &  \begin{tabular}[c]{@{}c@{}}Ubuntu 18.04.2 LTS OS, Intel Xeon{\textsuperscript \textregistered} Scalable 5118 \\ 12x2 Cores @2.3GHz, Samsung 256 GB DDR4 RAM,\\ 512x2 GB and 1.6x4 TB SSD, Mellanox 100G SmartX \\ NIC (2 ports), 16x6 GB Tesla T4 Nvidia{\textsuperscript \textregistered} GPUs  \end{tabular} \\ \hline
 
   Edge $\mu$-box  & \begin{tabular}[c]{@{}c@{}}Ubuntu 18.04.2 LTS OS, Supermicro SuperServer \\E300-8D (Mini-1U Server) with 4 @2.2GHz Intel\\Cores, 32 GB Memory, 240 GB HDD, 2x10 Gbps  \\+ 6x1 Gbps network interfaces   \end{tabular}  \\ \hline

\end{tabular}}
\label{specification}
\vspace{-10pt}
\end{table}
}
In this section, we discuss the methodology and experimental results obtained on the real network traffic dataset collected from the edge $\mu$-boxes to analyze and evaluate the proposed prediction model and validate its effectiveness in terms of two performance metrics (RMSE and R\textsuperscript{2}).

We collected multivariate network flow data at the visibility center for one month with an interval gap of 5-minutes using eBPF-based packet tracing software (Sec.~\ref{sec:dataset}). The collected dataset comprised 43000 time-series records with five features representing statistical features of data flow, divided into training (65\%) and test/validation (35\%). To train the prediction model, we deployed the Kubeflow and trained the LSTM-based encoder-decoder model on Jupyter Notebook with DL libraries (Tensorflow \& Keras) and Scikit-learn library on the accessed dashboard of deployed Kubeflow (c.f.~Fig.~\ref{fig:AIservice}). This setup enabled us to run the DL workloads on a fully automated and scalable cloud-native environment.  GPU resources of Intelligence Centers are used to run the deep learning jobs. Table.~\ref{specification} shows the hardware specifications of used resources for this experiment. 
\begin{figure}
\centering
\includegraphics[width=0.85\linewidth]{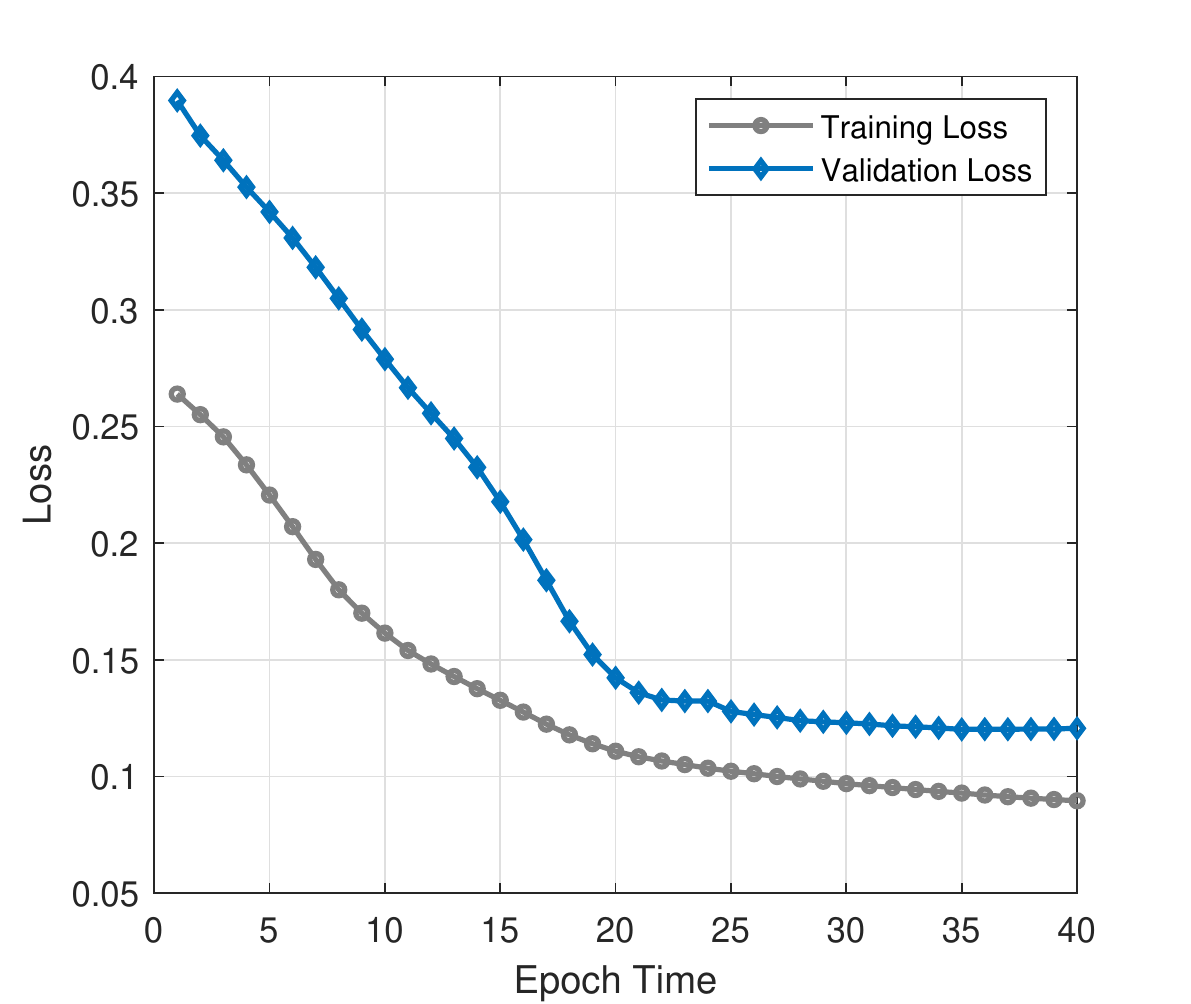}
\vspace{-7pt}
	\caption{Training loss and validation loss curves for our proposed DL-based prediction model.}
	\label{fig:loss}
	\vspace{-10pt}
\end{figure}
\begin{figure}
\centering
\includegraphics[width=0.85\linewidth]{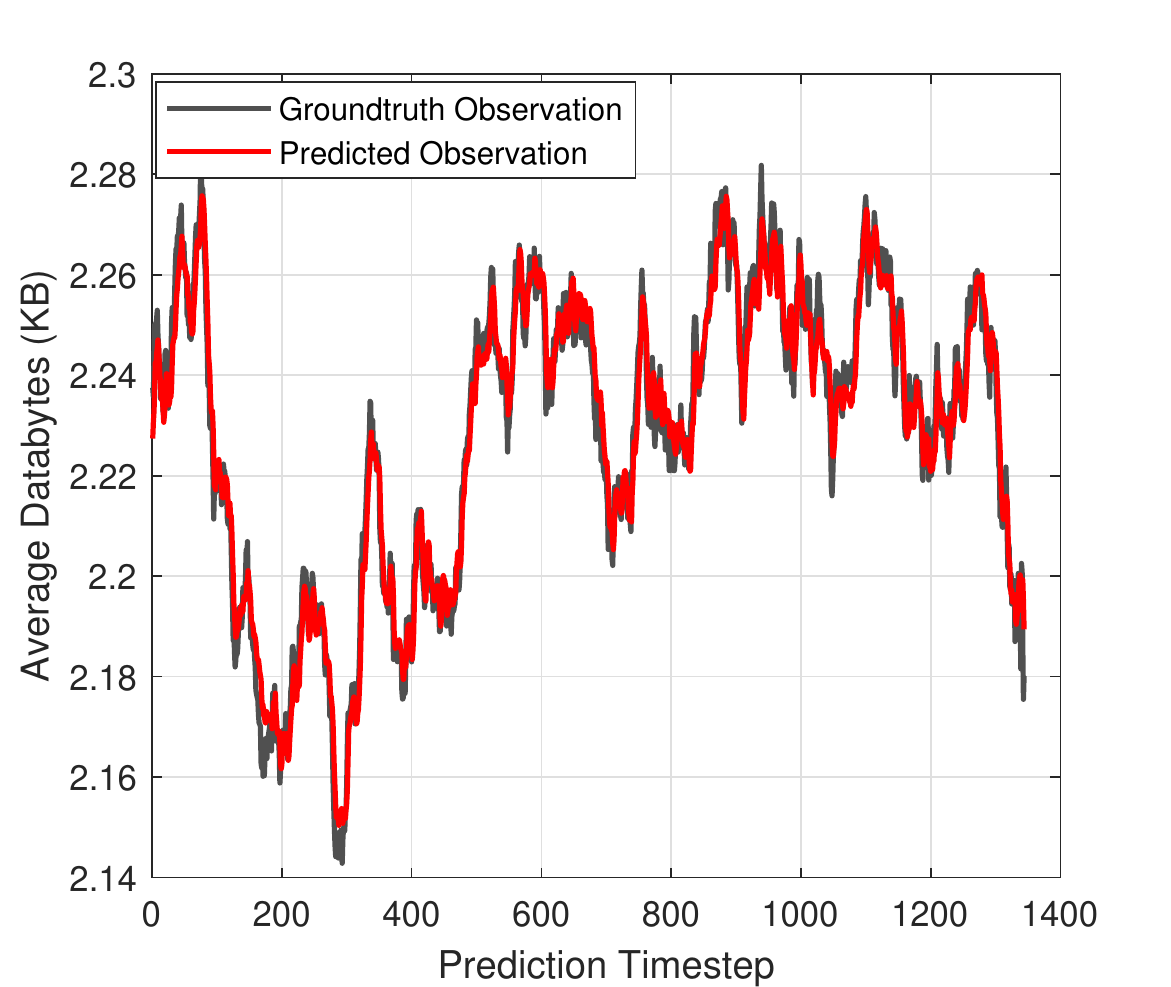}
\vspace{-7pt}
	\caption{Average databytes in data traffic predicted against the collected groundtruth observations.}
	\label{fig:average}
	\vspace{-10pt}
\end{figure}
A single-layer encoder and decoder architecture with time distributed layer is implemented with 100 LSTM cells in each layer. We applied the min-max function in the Scikit-learn library to normalize the value of the observed features in the range of $[-1,1]$. We trained the model on the past 20 hours of observation to predict the next 10 hours of output samples, compared with the groundtruth observations for model validation. Hyperparameters like batch size and epoch time for learning are kept fixed at 32 and 40, respectively. We selected the Adam optimizer to learn the optimized parameters while minimizing the $L_i^j$ in \eqref{eq7} during the training process. We applied a callback utility function "Learning Rate Schedule" to obtain the updated learning rate value through training from the defined range of $[1e^{-3},0.90\,e^{\textrm{epoch}}]$. It uses the updated learning rate on the Adam optimizer with the current epoch and current learning rate. Fig.~\ref{fig:loss} shows the trend of loss function during the training and validation stage of prediction mode against the epoch time. It reveals that beyond the epoch interval of 20, the validation and training loss is comparatively low and converging to avoid overfitting or underfitting in the prediction model.  
\begin{figure}
\centering
\includegraphics[width=0.85\linewidth]{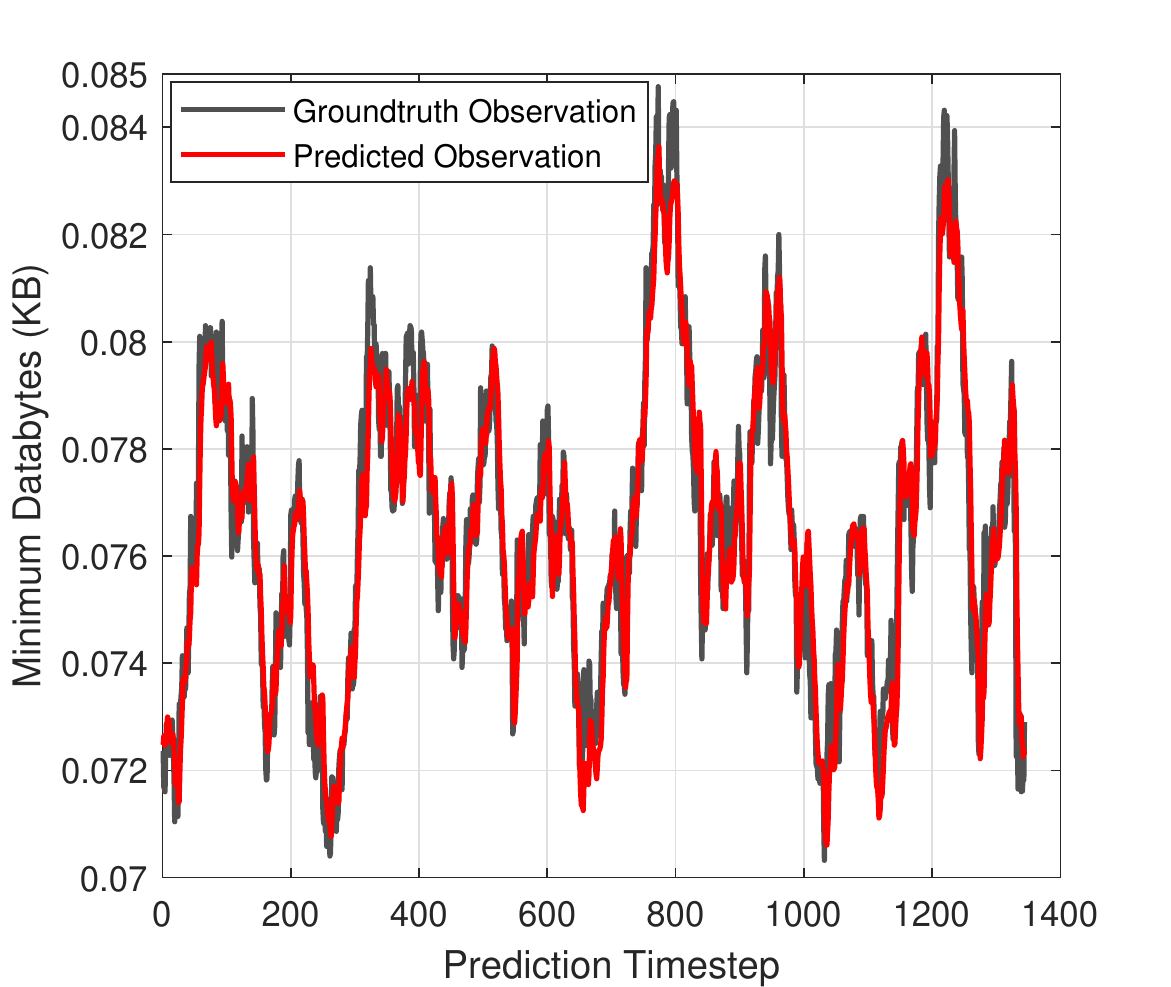}
\vspace{-7pt}
	\caption{Predicted minimum databytes in data traffic against the collected groundtruth observations.}
	\label{fig:minimum}
	\vspace{-10pt}
\end{figure}
\begin{figure}
\centering
\includegraphics[width=0.85\linewidth]{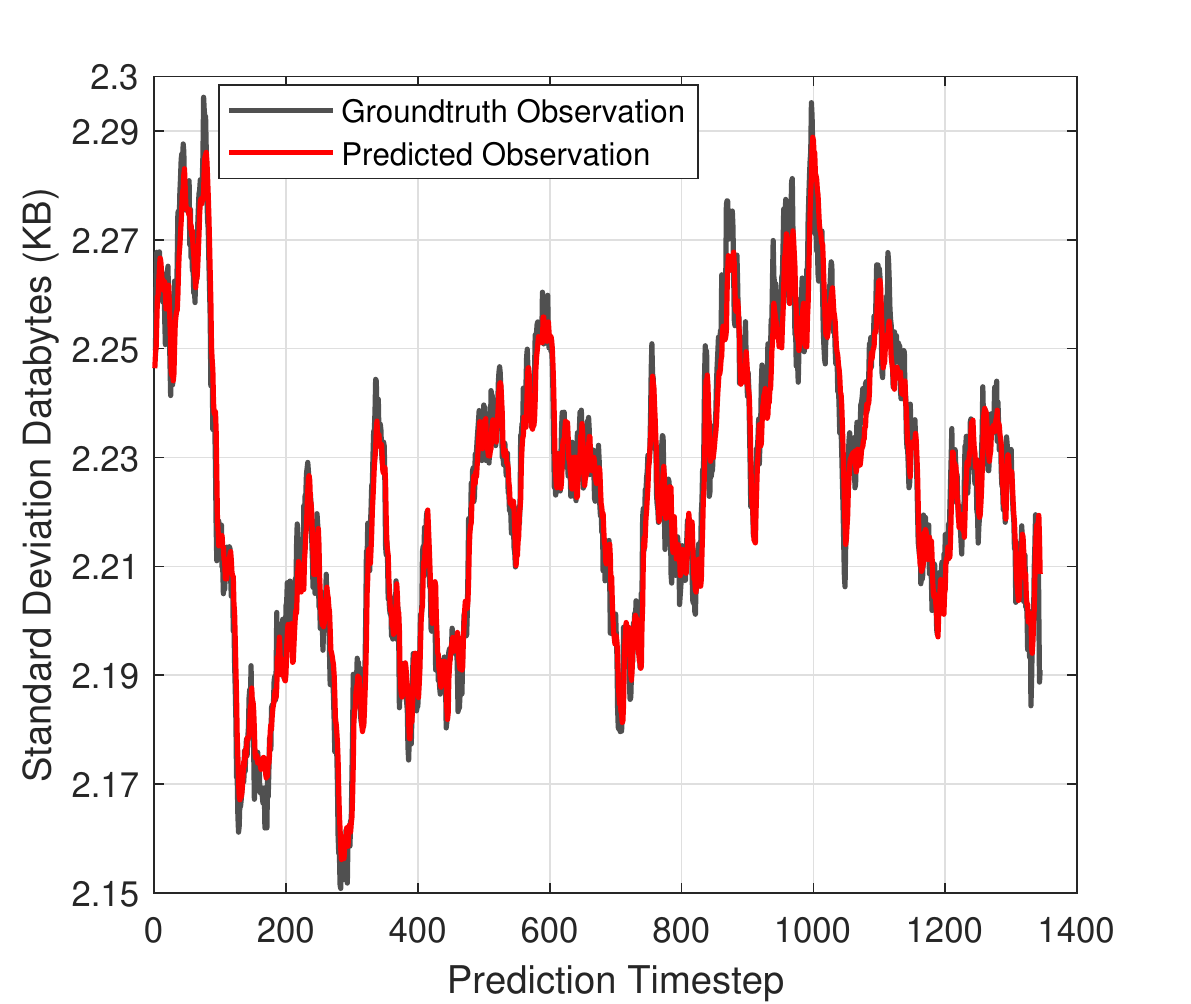}
\vspace{-7pt}
	\caption{Predicted standard deviation in databytes for data traffic predicted against the collected groundtruth observations.}
	\label{fig:std}
	\vspace{-10pt}
\end{figure}

The trained prediction model predicts the future samples of each statistical feature of network packet traffic in the selected future horizon window of 10 hours. Each predicted feature sample is verified against the ground truth observation. 
Fig.~\ref{fig:average} presents the statistics of average (mean) databytes recorded in the network flow at the edge $\mu$-boxes against the predicted average databytes by our learning model. Similarly, Fig.~\ref{fig:minimum} and Fig.~\ref{fig:std} show the predicted statistics of minimum (min) and standard deviation (std) in observed databytes of the network packet flow against the ground truth observations. 
These results show that our learning model can accurately learn and predict the future trend in mean, min, and std statistics of databytes at $\mu$-boxes over time and matches the recorded ground truth observations. 

\begin{figure}
\centering
\includegraphics[width=0.85\linewidth]{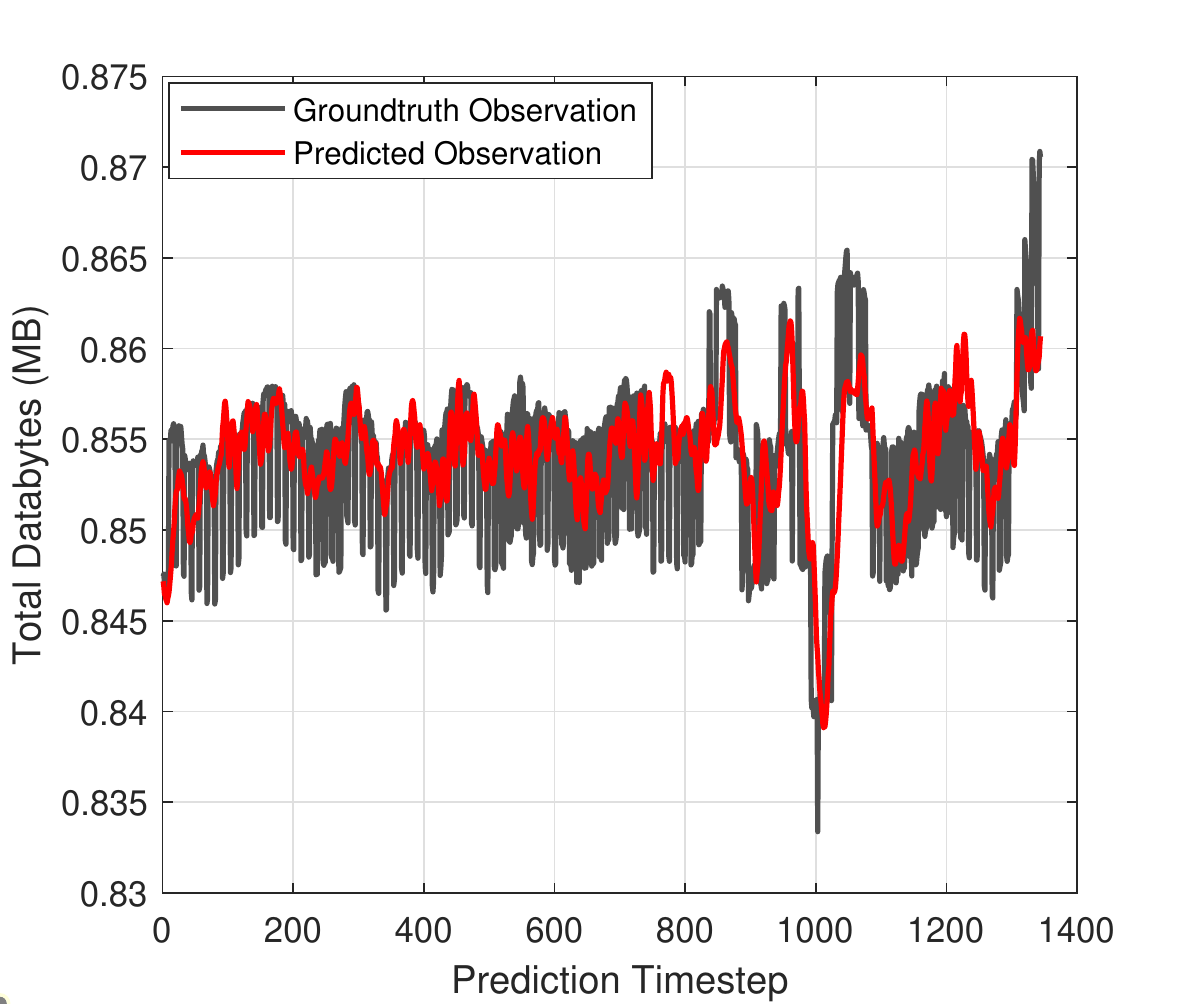}
\vspace{-10pt}
	\caption{Total databytes in data traffic predicted against the collected groundtruth observations.}
	\label{fig:total}
	\vspace{-12pt}
\end{figure}

Lastly, Fig.~\ref{fig:total} shows the predicted total traffic of the packet flows at the edge devices at the edge layer against the observed ground truth. It learns the total databytes statistics trend, which depicts the network traffic load over a specific period of time. Note that the statistics of avg, std, min, and maximum (max) databytes are recorded in kilobytes (KB) while total databytes are in megabytes (MB) units. Collectively predicting these five features characterizes the network traffic trend and accurately gives insight into the traffic statistics at the $\mu$-boxes of the edge layer. 

To analyze and validate the deviations between the learned model prediction samples and ground truth observations, we evaluated two performance metrics, RMSE and R\textsuperscript{2}, of each feature of data traffic. RMSE can take values from a range of $[0,\infty]$, while R2 takes a value in the range of $[0,1]$.
Table.~\ref{performancemetrics} shows the performance metrics for each data feature, showing that most of the features RMSE is closer to zero while R2 values are closer to 1. Low values of RMSE and R2 value closer to 1 imply that the learned model can accurately predict the five multivariate statistical features of data traffic. It can be observed that the performance of our model in predicting avg, min, max, and std statistics is better than total bytes as RMSE and R\textsuperscript{2} score of four features is reasonably good compared to the total databytes. 
\vspace{-5pt}

\section{Conclusion}
\label{sec:Conclusion}
Emerging computing techniques, AI, and state-of-the-art communication network enablers (SDN/NFV) are critical parts of the 6G vision, increasing the importance of intelligent network management. 
We proposed a DL-based novel intelligent prognosis technique for predicting statistical properties of data traffic incurred at the edge devices of the network. 
For this purpose, we captured, collected, and pre-processed the live TS traffic data from the edge $\mu$-boxes using the testbed network resources at GIST PG.
We orchestrated the Kubeflow deployment using K8s master at the orchestration center to train the LSTM-based seq2seq DL model on the collected TS data. We predicted various features of data traffic based on past observation into the future horizon window of 10 hours. We evaluated the predicted future observations with ground truth observation in terms of RMSE and R\textsuperscript{2}. Results showed that our model accurately predicts the future observations of all features.
For future work, network resource automation and scaling based on predicted traffic can be explored. 
{\renewcommand{\arraystretch}{1.3}
\begin{table}[t!]
\centering
	\caption{Prediction Performance for various features of Data Traffic} 
\scalebox{0.9}{	
	\begin{tabular}{|c|c|c|c|c|c|}
\hline
\textbf{\begin{tabular}[c]{@{}c@{}}Data \\ Feature\end{tabular}} & \textbf{\begin{tabular}[c]{@{}c@{}}Average\\ Databytes\end{tabular}} & \textbf{\begin{tabular}[c]{@{}c@{}}Min\\ Databytes\end{tabular}} & \textbf{\begin{tabular}[c]{@{}c@{}}Std\\ Databytes\end{tabular}} & \textbf{\begin{tabular}[c]{@{}c@{}}Total\\ Databytes\end{tabular}} & \textbf{\begin{tabular}[c]{@{}c@{}}Max\\ Databytes\end{tabular}}  \\ \hline
RMSE  &   5.33  & 8.63  &  6.03  & 231.64 & 33.12  \\ \hline
R\textsuperscript{2}   &  0.968    &   0.909   &   0.954  & 0.686 & 0.946   \\ \hline
\end{tabular}
}
  \label{performancemetrics}
  \vspace{-15pt}
\end{table} 
}

\end{document}